\documentclass{aa}
\usepackage{graphicx}
\begin{document}

\title{
UVES spectra of young brown dwarfs in Cha\,I: 
radial and rotational velocities 
  \thanks{Based on observations obtained at the European Southern
	Observatory at Paranal, Chile in program 65.L-0629 and 65.I-0011.}}


\author{V. Joergens \inst{1}
        \and E. Guenther \inst{2}
          }

   \institute{Max-Planck-Institut f\"ur Extraterrestrische Physik,
              Giessenbachstrasse 1, D-85748 Garching, Germany\\
              email: viki@mpe.mpg.de
         \and
              Th\"uringer Landessternwarte Tautenburg,
              Karl-Schwarzschild-Observatorium,
              Sternwarte 5,
              D-07778 Tautenburg,
              Germany}

   \offprints{V. Joergens}

   \date{Received; accepted}

\titlerunning{}
\authorrunning{Joergens \& Guenther}

   \abstract{Based on high-resolution UVES spectra we found that
		the radial velocity (RV) dispersion of nine of 
		twelve known young bona fide and candidate brown dwarfs 
		in the Cha\,I dark cloud is 2.0\,km\,s$^{-1}$, i.e.
		significantly smaller than
		the RV dispersion of T~Tauri stars 
		in Cha\,I (3.6\,km/s) and only slightly 
		larger than the dispersion 
		of the surrounding molecular
		gas (1.2\,km/s) (Mizuno et al. 1999).
		This result indicates that the majority of these brown dwarfs
		are not ejected with high velocity out of a dense
	region as proposed by some formation scenarios for brown dwarfs.
		The mean RV values are consistent with
		the objects being kinematic members of Cha\,I.
		The RV dispersion of the T~Tauri stars confined to the Cha\,I
		region is based on a
		compilation of T~Tauri stars 
		with known RVs from the literature plus 
		three T~Tauri stars 
		observed with UVES and unpublished RVs for nine T~Tauri stars.
		Time--resolved spectroscopy revealed RV variations for 
		five out of nine of the bona fide and candidate 
		brown dwarfs in Cha\,I, 
	which could be due to orbiting planets or surface features.
		Furthermore we derived rotational velocities $v\,\sin i$
		and the Lithium 6708\,{\AA} equivalent width.
        }

   \maketitle

%

\section{Introduction}
It is still an open
question how brown dwarfs form. 
They may form like planets in 
disks around normal stars. 
The 'brown dwarf desert' is not supporting this theory
unless the brown dwarfs are ejected after their formation
and can now be detected as freely floating brown dwarfs.
Sterzik \& Durison (1999) proposed the ejection
of brown dwarfs from cloud cores by three-body encounters.

Brown dwarfs may also form like stars by collapse of a cloud but
do not become stars because they formed out of relatively small cores.
There are several observational hints to significant circumstellar 
material around young brown dwarfs (e.g. Comer\'on et al. 2000 
for Cha\,I) supporting a star-like formation.
Recently Reipurth \& Clarke (2001) proposed that brown dwarfs
form by cloud fragmentation but failed to become 
stars because they have been ejected in the early accretion phase. 

Understanding the formation of brown dwarfs is
important because the number of brown dwarfs is at least
equal to the number of stars (Reid et al. 1999).
The ejection of a (proto-) brown dwarf
in the early accretion phase or later might have left observable signs. 
In this paper we report on 
high-resolution spectroscopy
of young bona fide and candidate brown dwarfs in Cha\,I,
which indicate that there is probably no run-away brown dwarf among them.

The Chamaeleon cloud complex is comprised of three main clouds.
Cha\,I is the most active star forming cloud among them and 
one of the most promising grounds for observational projects on
very low-mass objects.
Recently twelve low-mass M6--M8--type objects,
Cha\,H$\alpha$\,1 to 12, have been detected in the center of Cha\,I
with ages in the range of 1 to 5\,Myrs
(Comer\'on et al. 1999, 2000).
Their masses are below or near the border line separating brown dwarfs
and very low-mass stars. 
Four of them are confirmed bona fide brown dwarfs 
($\sim$30 to 50\,M$_{\mbox{\tiny Jup}}$)
(Neuh\"auser \& Comer\'on 1998, 1999; Comer\'on et al. 2000).

\section{Spectroscopy}
Using the high-resolution Echelle spectrograph UVES 
(Dekker\,et\,al.\,2000)
at the VLT, we carried out
spectroscopic observations of nine
bona fide and candidate brown dwarfs (M6-M8) 
(Cha\,H$\alpha$\,1 to 8 and 12)
and three mid-M-type T~Tauri stars (B\,34, CHXR\,74 and Sz\,23) 
in Cha\,I between March and May 2000.
For each object at least
two spectra separated by a few weeks have been obtained.

UVES is ideal for high precision measurements of 
RVs of faint objects.
The spectra have been taken using the red arm of the two-armed UVES 
spectrograph equipped with a mosaic of two CCDs (blue side of red arm: EEV; 
red side of red arm: MIT-LL).
The wavelength coverage is 6600\,{\AA} to 10400\,{\AA} and 
the spectral resolution $\rm \lambda / \Delta \lambda=40000$. 
The spectra have been taken with a slit of 1$\arcsec$ to 1.2$\arcsec$ 
during a seeing of 0.4$\arcsec$ to 1.0$\arcsec$.

The spectra have been optimally extracted, including bias correction,
flat fielding, cosmic ray elimination and sky subtraction using the Echelle 
package of IRAF\footnote{IRAF is distributed by the National Optical
   Astronomy Observatories,
   which is operated by the Association of Universities for Research in
   Astronomy, Inc. (AURA) under cooperative agreement with the National
   Science Foundation.}. 

\section{Radial velocities}
%
\begin{figure}[t]
\begin{center}
\includegraphics[height=.3\textwidth,angle=270]{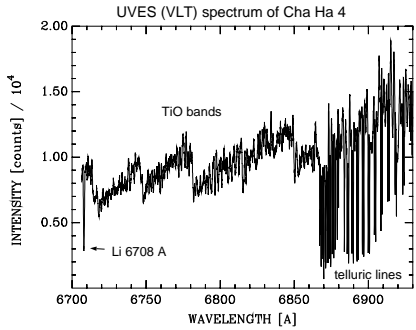}
\end{center}
\caption[]{\label{spec} Small part of an UVES Echelle spectrum of 
Cha\,H$\alpha$\,4.
}
\end{figure}
For the determination of precise RVs 
we used the telluric O$_2$ lines 
as wavelength reference. They
are produced in the Earth atmosphere and 
show up in the red part of the optical spectral range
(cp. Fig.\,\ref{spec}). They 
are stable up to $\sim$\,20\,m\,s$^{-1}$ 
(Balthasar et al. 1982, Caccin et al. 1985).

Heliocentric RVs are determined by 
cross-correlating plenty of stellar lines of the 
object spectra against a template spectrum and 
locating the correlation maximum.
A mean UVES spectrum of CHXR\,74 served as 
template. The zero point of its velocity has been 
determined by means of the Lithium line at 6708\,{\AA}.

For the measurements of Doppler shifts of stellar features
we carefully selected appropriate wavelength regions, which are
not affected by telluric lines, cosmic defects of the CCD or fringes of the 
CCD in the near IR.
We achieved a RV precision of $\sim$200\,m\,s$^{-1}$ for a S/N of $\sim$20
in agreement with the expectations for this S/N (Hatzes \& Cochran 1992).
The precision of the RVs is limited by the S/N of the spectra and not by
systematic effects. 

The mean heliocentric RVs are given in Table\,\ref{tab}.
They are consistent with
RVs measured by Neuh\"auser \& Comer\'on (1999) within the
measurements uncertainties for 
Cha\,H$\alpha$\,1, 3, 4, 5, 7, 
B\,34 and CHXR\,74.
However, the values for 
Cha\,H$\alpha$\,8 and Sz\,23
are discrepant 
by more than 1\,$\sigma$ and the RVs for
Cha\,H$\alpha$\,2 and 6 by 2\,$\sigma$.
This may be a hint to long-term single-lined spectroscopic binaries.

Due to a low S/N for Cha\,H$\alpha$\,7 
only a mean RV and an upper limit of $v\,\sin i$ and the Lithium
equivalent width could be determined.

\section{Kinematics of brown dwarfs and T~Tauri stars in the Cha\,I cloud}
\begin{figure}[t]
\begin{center}
\includegraphics[width=.29\textwidth,angle=270]{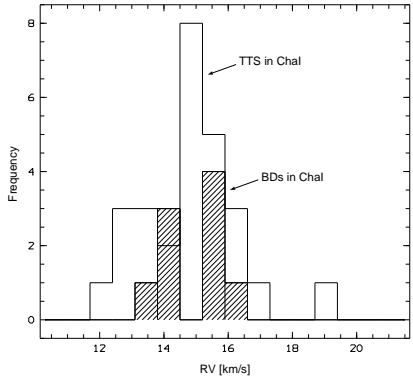}
\end{center}
\caption[]{
Histogram of mean RVs of nine bona fide
and candidate brown dwarfs (hashed) and
for 27 T~Tauri stars in Cha\,I.
}
\label{hist}
\end{figure}

%
\begin{table}[t]
\begin{center}
\caption{\label{tab} Parameters derived from UVES spectra:
mean heliocentric radial velocity RV in [km\,s$^{-1}]$, projected 
rotational velocity $v\,\sin i$ and
equivalent width EW of Lithium (6708\,{\AA}).
Last columns: (sub)stellar radii and 
upper limits for rotational periods.
}
\begin{tabular}{lccccc}
\hline
object& RV & $v\,\sin i$ & EW(Li) & R$_{\star}$   & P$_{\rm max}$\\
& $\pm$0.2 &[km\,s$^{-1}]$ &[\AA] & [R$_{\odot}$] & [d]\\
\hline
Cha\,H$\alpha$\,1   & 15.5 &  7.6$\pm$2.2 & 0.14 $\pm$0.10 &0.46& 3.1 \\ 
Cha\,H$\alpha$\,2   & 15.4 & 12.8$\pm$1.2 & 0.30 $\pm$0.08 &0.73& 2.9 \\
Cha\,H$\alpha$\,3   & 14.2 & 21.0$\pm$1.6 & 0.26 $\pm$0.08 &0.77& 1.9 \\
Cha\,H$\alpha$\,4   & 15.4 & 18.0$\pm$2.3 & 0.46 $\pm$0.11 &0.89& 2.5 \\
Cha\,H$\alpha$\,5   & 15.5 & 15.4$\pm$1.8 & 0.44 $\pm$0.10 &0.83& 2.7 \\
Cha\,H$\alpha$\,6   & 16.1 & 13.0$\pm$2.8 & 0.32 $\pm$0.08 &0.68& 2.6 \\
Cha\,H$\alpha$\,7   & 13.7 & $\leq 10 $   & $\leq$ 0.09    &0.37& $\geq 1.9$\\
Cha\,H$\alpha$\,8   & 14.5 & 15.5$\pm$2.6 & 0.33 $\pm$0.15 &0.59& 1.9\\
Cha\,H$\alpha$\,12  & 13.8 & 25.7$\pm$2.6 & 0.37 $\pm$0.10 &0.66& 1.3\\
\hline
B\,34               & 16.5 & 15.2$\pm$1.9 & 0.60 $\pm$0.13 &\\ 
CHXR\,74            & 15.1 & 14.1$\pm$1.6 & 0.63 $\pm$0.04 &\\ 
Sz\,23              & 15.8 & 17.3$\pm$3.4 & 0.31 $\pm$0.08 &\\ 
\hline
\end{tabular}
\end{center}
\end{table} 
%
Neuh\"auser \& Comer\'on (1999) determined a 
mean 
RV of $\sim$14.6\,km\,s$^{-1}$ and a total range of 11\,km\,s$^{-1}$
for Cha\,H$\alpha$\,1 to 8
from medium resolution spectra.
The measurements of precise RVs for  
Cha\,H$\alpha$\,1 to 8 and Cha\,H$\alpha$\,12 with UVES 
allow us to study the kinematics of these bona fide and candidate 
brown dwarfs with high accuracy. 
We find that their RVs lie close together, 
only spanning a range of 2.4\,km\,s$^{-1}$.
The mean RV is 14.9\,km\,s$^{-1}$ and
the RV dispersion is 2.0\,km\,s$^{-1}$ 
(cp. Table\,\ref{tab} and Fig.\,\ref{hist}).

The Cha\,H$\alpha$ objects are located at the periphery of one
of the six cloud cores (No.\,5) in Cha\,I in a region with a relatively
high density of young stellar objects. 
The mean RV of the molecular gas of the Cha\,I cloud
and also of the cloud core No.\,5 is 15.3\,km\,s$^{-1}$ (Mizuno et al. 1999). 
The mean RV of the studied brown dwarfs is consistent with
this velocity of the gas and therefore with
the objects being kinematic members of Cha\,I.
Mizuno et al. determined the RV dispersion
of the gas of core No.\,5 to 1.2\,km\,s$^{-1}$.
The brown dwarfs show a slightly larger 
RV dispersion (2.0\,km\,s$^{-1}$) than the surrounding molecular gas
but basically reflect the motion of the gas.

The relatively small RV dispersion of the studied bona fide 
and candidate brown dwarfs gives suggestive evidence that there 
is no run-away brown dwarf among them.
We cannot rule out that some of them have a larger space velocity dispersion 
since RVs are tracing only space motions in one dimension.
Nevertheless our finding indicates that the majority of the 
nine Cha\,H$\alpha$ objects are not ejected with high velocities
out of a dense region as proposed in formation scenarios 
(Sterzik \& Durisen 1999, Reipurth \& Clarke 2001).
Some or all of the brown dwarfs may still
have been 'ejected' with less than escape velocity into an extended orbit
around another component of a multiple system.
None of the studied brown dwarfs
is closer to a known T~Tauri star than 4600\,AU,
i.e. it is unlikely that one of them 
is still bound to a star. There is still the possibility
that the parent star itself was later 
ejected with escape velocity and left an unbound brown dwarf.
%
\begin{table}[t]
\begin{center}
\caption{\label{tts} List of T~Tauri stars in Cha\,I with known RVs
with a precision of 2\,km\,s$^{-1}$ or better from:
$^1$ Dubath et al. (1996), $^2$ Covino et al. (1997), 
$^3$ Guenther et al.,in prep. and from UVES observations 
(B\,34, CHXR\,74, Sz\,23: see Table\,\ref{tab}).}
\begin{tabular}{lcc}
\hline
object & RV [km\,s$^{-1}$]    & $v\,\sin i$ [km\,s$^{-1}$] \\ 
\hline
Sz\,4$^1$  & 16.5 $\pm$1.3 & 16.6 $\pm$3.3 \\
Sz\,6$^1$  & 14.9 $\pm$0.8 & 38.0 $\pm$1.5\\
Sz\,9$^1$  & 14.7 $\pm$0.3 & 15.2 $\pm$1.5\\
Sz\,11$^1$ & 15.1 $\pm$0.5 & 16.2 $\pm$2.3\\
Sz\,15$^1$ & 17.2 $\pm$0.6 &  4.9 $\pm<$7.9\\
Sz\,19$^1$ & 13.5 $\pm$0.6 & 35.7 $\pm$1.1\\
Sz\,20$^1$ & 15.4 $\pm$1.3 & 23.9 $\pm$2.7\\
Sz\,36$^1$ & 12.9 $\pm$0.9 & 7.5 (+2.7-4.8)\\
Sz\,41$^1$ & 13.9 $\pm$0.4 & 34.4 $\pm$1.4\\
Sz\,42$^1$ & 15.1 $\pm$0.3 & 27.2 $\pm$1.3 \\
RXJ1109.4-7627$^2$   & 13.1 $\pm$2.0 & 14.5 $\pm$3 \\
B\,33$^2$ (CHXR\,25) & 13.0 $\pm$2.0 &  --       \\
F\,34$^2$          & 14.0 $\pm$2.0 & 55 $\pm$3   \\
RXJ1111.7-7620$^2$ & 19.0 $\pm$2.0 & 23 $\pm$3   \\
RXJ1112.7-7637$^2$ & 16.0 $\pm$2.0 & 11 $\pm$3   \\
CS Cha$^3$         & 14.9          &\\
CT Cha$^3$         & 15.5 $\pm$1.4 & \\
CV Cha$^3$         & 15.6 $\pm$0.9 &\\
SX Cha$^3$         & 13.4 $\pm$0.9 &\\
SY Cha$^3$         & 12.7 $\pm$0.1 &\\
TW Cha$^3$         & 15.7 $\pm$1.2 &\\
VW Cha$^3$         & 15.1 $\pm$0.1 &\\
VZ Cha$^3$         & 14.7          &\\
WY Cha$^3$         & 12.1          &\\
\hline
\end{tabular}
\end{center}
\end{table}

We compared the RV distribution of the 
bona fide and candidate brown dwarfs also with those of
T~Tauri stars.
Radio observations by Mizuno et al. (1999)
revealed that the three main clouds in the Chamaeleon star forming 
region differ to a large extent in their star formation properties
and also the RVs of the molecular gas 
vary between the clouds (differences up to 3.6\,km$^{-1}$), 
whereas they are relatively constant 
($\sim$1\,km\,s$^{-1}$) within each single cloud.
Therefore it is reasonable to compare the kinematics of the
brown dwarfs in Cha\,I with those of T~Tauri stars also
confined to the Cha\,I star forming cloud.

RVs of T~Tauri stars in Cha\,I have been
measured by Dubath et al. (1996), Covino et al. (1997), 
Neuh\"auser \& Comer\'on (1999) and by us 
and are listed in Table\,\ref{tts}.
Furthermore unpublished RVs of T~Tauri stars based on
FEROS spectra have been included (Guenther et al., in prep.).
The T~Tauri stars have RVs in the range of 
[12.1$\dots$19.0\,km\,s$^{-1}$] with 
a mean RV of 14.9\,km\,s$^{-1}$ and a dispersion of 
3.6\,km\,s$^{-1}$ (cp. Fig.\,\ref{hist}). 

The mean RV of the T~Tauri stars matches very well the ones of the
bona fide and candidate brown dwarfs in Cha\,I, 
whereas the dispersion as well as the total range of RVs of the
T~Tauri stars is significantly larger than the ones of the brown dwarfs.
The stellar activity of T~Tauri stars probably account
for this discrepancy since it has been shown that 
T~Tauri stars exhibit a 'RV noise' of the order of $\sim$2\,km\,s$^{-1}$ due
to stellar activity (Guenther et al. 2000).
Moreover it may also play a role that 
the brown dwarfs and brown dwarf candidates are all situated
in a small area at the periphery of one cloud core whereas the 
T~Tauri stars are distributed over the whole Cha\,I region.
The RVs of the six cloud cores within Cha\,I 
differ by $\pm$0.25\,km\,s$^{-1}$ (Mizuno et al. 1999). 

%
%
%
\section{Radial velocity variations}
The analysis of UVES spectra taken at different times yielded to the 
detection of significant RV variations
for five out of nine bona fide and candidate 
brown dwarfs in Cha\,I. 
They could be caused by motion of the brown dwarf due to an orbiting 
companion or by 
magnetically induced surface features.
%
%
The detected RV variations are of the order of 1\,km\,s$^{-1}$. 
If they are caused by an orbiting object this 
would have a mass of a few Jupiter masses depending on the 
orbital parameters (Joergens et al. 2001).
%
%
More details will be given later after additional follow-up
observations.

\section{Projected rotational velocities $v\,\sin i$}
\label{vsinikapitel}

\begin{figure}[t]
\begin{center}
\includegraphics[width=.3\textwidth,angle=270]{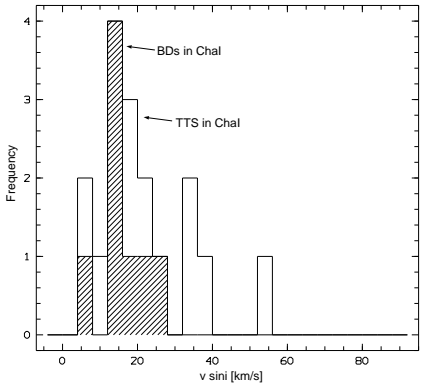}
\end{center}
\caption[]{\label{vsinihist}
Histogram of 
$v\,\sin i$ for brown dwarfs (hashed) and T~Tauri stars
in Cha\,I. For details see Sect.\,\ref{vsinikapitel}.
}
\end{figure}

Projected rotational velocities have been measured
using the telluric lines for determining the instrumental profile of
the spectrograph and assuming a solar-like center-to-limb variation.
We derived $v\,\sin i$ values in the range from 
8\,km\,s$^{-1}$ to 26\,km\,s$^{-1}$ (Table \ref{tab}).
Fig.\,\ref{vsinihist} shows the $v\,sin i$ distribution of the 
brown dwarfs in comparison with that of T~Tauri stars 
listed in Table\,\ref{tts}.
Both distributions peak at a $v\,\sin i$ of 
17 to 21\,km\,s$^{-1}$,
indicating that there is no crucial difference between the 
rotational velocities of the studied brown dwarfs and T~Tauri stars.

Based on $v\,\sin i$ and the radius of the object an upper limit of the 
rotational period can be derived:
\begin{equation}
P_{\rm max}[\mbox{d}]= 50.6145 \frac{R [\mbox{R}_{\odot}]}{(v \sin i) [\mbox{km}\,\mbox{s}^{-1}]}
\end{equation}
The radii of the objects are estimated from
bolometric luminosities 
and effective temperatures given by Comer\'on et al. (2000).
The approximate upper limits for the rotational periods
are between one and three days (Table\,\ref{tab}). 

\section{Lithium equivalent width}

All brown dwarfs and low-mass stars observed with UVES
show lithium absorption at 6708\,{\AA}. The measured 
equivalent width is given in Table\,\ref{tab}
for all objects. For Cha\,H$\alpha$\,7 
only an upper limit of EW(Li) was determined due to a low S/N.
For eight of the twelve bona fide and candidate brown dwarfs in
Cha\,I Lithium detection and equivalent width measurements have 
been reported by Neuh\"auser \& Comer\'on (1999).
Our UVES spectra allow us to add 
the Lithium detection of Cha\,H$\alpha$\,12.

\section{Summary}

Based on high-resolution UVES spectra of bona fide and candidate
brown dwarfs and of T~Tauri stars in Cha\,I we determined
RVs with a precision of $\sim$200\,m\,s$^{-1}$.

We found that the RV dispersion of nine of the twelve
bona fide and candidate brown dwarfs in Cha\,I is
2.0\,km\,s$^{-1}$, i.e.
significantly smaller than the RV dispersion of T~Tauri stars (3.6\,km/s) 
in this cloud
and slightly larger than the dispersion of the surrounding molecular
gas (1.2\,km/s) (Mizuno et al. 1999). 
This result indicates that the majority of the
bona fide and candidate brown dwarfs in Cha\,I are
not ejected with high velocities out of a dense region
as proposed in some formation scenarios 
(Sterzik \& Durisen 1999, Reipurth \& Clarke 2001).
Some or all of the brown dwarfs may still
have been thrown with less than escape velocity into an extended orbit
around another component of a multiple system. 

The kinematic study of the T~Tauri stars in Cha\,I
was based on a compilation of all T~Tauri stars located in Cha\,I
where RVs were known to better than 2\,km/s
(including our UVES data for three T~Tauri stars
as well as unpublished data taken with FEROS).
The mean RV of the T~Tauri stars is 14.9\,km/s, i.e. the 
same as for the brown dwarfs. The larger RV dispersion 
of the T~Tauri stars can at least partly be attributed
to 'RV noise' caused by stellar activity (Guenther et al. 2000).

Time-resolved spectroscopy revealed significant RV variations
for five of the bona fide and candidate brown dwarfs in Cha\,I, 
which may be caused by orbiting planets or spots on the surface
(Joergens et al. 2001). 

Determination of $v\,sin i$ showed that the brown dwarfs 
do not rotate crucially faster than
the T~Tauri stars in the same cloud.
The $v\,sin i$ values together with radii derived from the 
literature constrained the maximum rotational periods for the individual 
brown dwarfs to one to three days.
Last not least we detected lithium in absorption for all
studied objects and measured the equivalent width -- 
for Cha\,H$\alpha$\,12 this is the first lithium detection. 

The data presented in this paper showed once more
that the brown dwarfs in Cha\,I
form a very homogenous sample and
are highly interesting 
astrophysical 
objects.

\begin{acknowledgements}
      We acknowledge helpful discussions on the topic of this paper with
      R. Neuh\"auser, R. Durisen, K. Tachihara and F. Comer\'on.
      Furthermore  we like to acknowledge the brilliant work of the 
      ESO staff at Paranal. 
      VJ acknowledges grant from the Deutsche Forschungsgemeinschaft
      (Schwerpunktprogramm `Physics of star formation').
\end{acknowledgements}

\end{document}